\documentclass[11pt,
showpacs,
preprintnumbers,amsmath,amssymb,
aps,prd,nofootinbib,eqsecnum,a4paper]{revtex4}
\usepackage{epsfig}
\usepackage{graphicx,epsf}
\usepackage{color}
\usepackage{bm}
\usepackage{psfrag}
\usepackage[symbol]{footmisc}
\usepackage{tensor}
\usepackage{hyperref}
\def\beq{\begin{equation}}
\def\eeq{\end{equation}}
\def\bea{\begin{eqnarray}}
\def\eea{\end{eqnarray}}
\def\bi{\begin{itemize}}
\def\ei{\end{itemize}}
\def\cs2{c_{\rm{s}}^2}
\def \beg {\begin{enumerate}}
\def \en {\end{enumerate}}

\def\M0{{\cal M}_0}

\DeclareMathOperator{\sech}{sech}

\begin{document}

\title{Viability of generalized $\alpha$-inflation from Planck, ACT, and DESI Data}

\author{Gabriel Germ\'an\footnote[1]{\href{mailto:gabriel@icf.unam.mx}{gabriel@icf.unam.mx}}, Juan Carlos Hidalgo \footnote[2]{\href{mailto:hidalgo@icf.unam.mx}{hidalgo@icf.unam.mx}}}

\affiliation{
$^{}$Instituto de Ciencias F\'{i}sicas, Universidad Nacional Aut\'{o}noma de M\'{e}xico,\\Av. Universidad S/N. Cuernavaca, Morelos, 62210, M\'{e}xico}
%

\begin{abstract}

We study inflationary constraints from reheating on two classes of single-field inflationary models: a generalized $\alpha$-attractor with potential $V(\phi) = V_0 \!\left(1 - \sech^p\!\left[\phi/(\sqrt{6\alpha}\,M_{Pl})\right]\right)$ and the $\alpha$-Starobinsky model with potential $V(\phi) = V_0 \!\left(1 - e^{- \sqrt{2/(3\alpha)}\, \phi / M_{Pl}} \right)^2$. Using a semi-analytical relation that connects inflationary dynamics to reheating, we solve for the horizon-crossing field value and calculate the scalar spectral index $n_s$, the tensor-to-scalar ratio $r$, and the reheating temperature $T_{re}$. We consider three perturbative decay channels of the inflaton: gravitational, Yukawa (fermionic), and scalar. The parameter space spanned by $(\alpha,p,\omega_{re})$ for the first model and $(\alpha,\omega_{re})$ for the second is explored and compared with recent measurements from Planck and ACT DR6, as well as BAO from DESI DR2.
For clarity, we emphasize that ACT DR6 alone is fully compatible with Planck regarding $n_s$. The upward shift toward $n_s \simeq 0.975$ appears only when DESI DR2 BAO are combined with CMB datasets (Planck+DESI, ACT+DESI, SPT-3G+DESI), and is likewise seen in the P--ACT--LB2+DESI combination. CMB-only combinations (e.g., Planck+ACT or Planck+ACT+SPT-3G) primarily refine constraints and do not by themselves raise $n_s$. Our comparisons and model assessments are therefore made with this distinction in mind: CMB-only constraints versus CMB+DESI combinations.
Within the $\sech^p$ model, decreasing $p$ shifts predictions toward the $n_s$ values preferred by the CMB+DESI combinations, but may yield too-large $r$; we find $p=1/4$ provides a viable compromise region. For the $\alpha$-Starobinsky case, increasing $\alpha$ raises $n_s$, but values $\alpha \gtrsim 10$ generally exceed observational bounds on $r$. The three decay channels modify the allowed ranges of $T_{re}$ and $n_s$, with the scalar channel giving the strongest restrictions. We conclude that both models remain consistent with current CMB-only data in restricted regions of parameter space, and that residual tension with the higher $n_s$ favored by CMB+DESI persists. 
\end{abstract}


\maketitle

\section{Introduction} \label{Intro}

Cosmological inflation (for reviews, see e.g.,~\cite{Linde:1984ir}-\cite{Martin:2013tda}) provides a framework for generating the primordial fluctuations that seed cosmic microwave background (CMB) anisotropies and large-scale structure. In single-field models, the inflationary dynamics and the subsequent reheating stage \cite{Bassett:2005xm}-\cite{Amin:2014eta} are directly connected, and their predictions can be constrained by measurements of the scalar spectral index $n_s$ and the tensor-to-scalar ratio $r$. The current data show a mild tension between CMB-only inferences and CMB+BAO combinations: Planck alone prefers $n_s \simeq 0.965$ \cite{Planck:2018jri}, while ACT DR6 alone is fully compatible with Planck \cite{ACT:2025fju,ACT:2025tim}. However, when either ACT DR6 or Planck is combined with DESI DR2 BAO \cite{DESI:2024uvr,DESI:2025zgx}, the joint fits favor $n_s \simeq 0.975$, a difference at the $\sim 2\sigma$ level. Including the recent SPT-3G release \cite{SPT-3G:2025bzu} further completes the CMB comparison; the same upward shift appears when SPT-3G is combined with DESI DR2, and likewise in the P–ACT–LB2 combination reported by ACT \cite{Ferreira:2025lrd,Ellis:2025ieh}. 

To be clear: the upward shift in the scalar spectral index $n_s$ arises only when DESI DR2 BAO data are combined with CMB experiments (Planck, ACT, or SPT). DESI does not measure $n_s$ directly; rather, under $\Lambda$CDM its BAO constraints on $(r_d h, \Omega_m)$ are in mild tension with the corresponding CMB constraints. Because $n_s$ is correlated with these parameters in CMB-only fits, the combined analysis leads to $n_s \simeq 0.975$. By contrast, CMB-only combinations such as Planck+ACT or Planck+ACT+SPT-3G do not drive an upward shift in $n_s$; adding SPT-3G in CMB-only datasets primarily refines constraints and improves control of small-scale systematics. The BAO–CMB tension therefore originates specifically from DESI DR2, not from SPT-3G. This behavior is also seen in the P–ACT–LB2 and SPT-3G combinations with DESI, which both exhibit the same upward shift in $n_s$ when DESI is included.

BAO provide a precise probe of the cosmic expansion history. The DESI DR2 survey maps the late-time BAO imprint in the 3D distribution of galaxies, quasars, and the Lyman-$\alpha$ forest ($0.2 \lesssim z \lesssim 4$). The relevant standard ruler is $r_d$, set by acoustic waves in the photon–baryon plasma prior to recombination. This scale first appears in the CMB as the spacing of the primordial acoustic peaks, and later reemerges in large-scale structure as baryon acoustic oscillations in the clustering of galaxies and other tracers. By contrast, SPT-3G constrains the same scale $r_d$ directly through high-resolution measurements of CMB temperature and polarization anisotropies at $z \sim 1100$, with additional information provided by CMB lensing up to $z \sim 5$. Thus, in combined analyses, DESI tracks the relative stretching of the BAO ruler since recombination, while CMB experiments (Planck, ACT, SPT-3G) define its absolute scale; the reported upward shift in $n_s$ emerges only when these two classes of measurements are analyzed jointly under $\Lambda$CDM.

Here we restrict to the datasets reported by the ACT DR6 collaboration, which motivates a careful test of inflationary models, taking into account the reheating phase. The primordial quantities inferred from the CMB (such as $n_s$ and $A_s$) depend on the expansion history between horizon exit and the end of reheating, so the mapping from model parameters to observables is sensitive to the reheating dynamics. Reheating is described by the effective equation-of-state parameter $\omega_{re}$, the reheating temperature $T_{re}$, and the number of $e$-folds during reheating $N_{re}$. In a perturbative treatment, the inflaton can decay, in particular, via gravitational, Yukawa, or scalar couplings. Although this approximation neglects non-linear processes, it provides a clear link between model parameters and predictions for $n_s$, $r$, and $T_{re}$.

A quantitative relation between inflation and reheating is given by \cite{Liddle:1994dx}-\cite{German:2025mzg}\footnote{The precise form presented here is derived in detail in \cite{German:2024rmn}.}
\begin{align}
\label{principal}
N_k = & \ln\left( \frac{2}{k_p} \left( \frac{43}{11 g_{s,re}} \right)^{1/3} \pi \sqrt{A_s}\, a_0 T_0 \right) 
+ \frac{1}{3(1 + \omega_{re})} \ln\left( \frac{g_{re}}{540 A_s} \right) \nonumber \\
&+ \frac{1 - 3 \omega_{re}}{3(1 + \omega_{re})} \ln \left( \frac{T_{re}}{M_{Pl}} \right)
+ \frac{1}{3(1 + \omega_{re})} \ln\left( \frac{\left( M_{Pl} f^{\prime}(\phi_k) \right)^{1 + 3 \omega_{re}}}{f(\phi_k)^{3 \omega_{re}} f(\phi_{end})} \right),
\end{align}
where $N_k = \ln(a_{end}/a_k)$ is the number of $e$-folds between horizon exit of the pivot mode $k_p=0.05\,\mathrm{Mpc}^{-1}$ and the end of inflation. Here $A_s$ is the amplitude of scalar perturbations, $a_0$ and $T_0$ are today’s scale factor and CMB temperature, and $g_{re}, g_{s,re}$ are the effective relativistic and entropy degrees of freedom at reheating. The potential is written as $V(\phi)=V_0 f(\phi)$, with $f(\phi)$ model-dependent; $f'(\phi)$ denotes its derivative with respect to $\phi$, and $\phi_{end}$ is defined by $\epsilon(\phi_{end})=1$. For $\omega_{re}=1/3$ the explicit dependence on $T_{re}$ cancels. Solving Eq.~(\ref{principal}) determines the inflaton value at horizon crossing $\phi_k$, from which $n_s$, $r$, and $T_{re}$ follow.

We study two models within this framework. The first is a generalized $\alpha$-attractor with a potential of the form (\ref{potsech}) \cite{German:2021rin}, which reduces to the standard case for $p=2$. The second is the $\alpha$-Starobinsky potential \cite{Ellis:2013nxa}, \cite{Kallosh:2013yoa}, \cite{Ellis:2019bmm}, which reproduces the Starobinsky model for $\alpha=1$. 
Both models are consistent with Planck data, but their agreement with the higher $n_s$ values obtained only in CMB+DESI combinations requires further study.

Our aim is to test these models under different reheating assumptions, compare their predictions with Planck and ACT+DESI data, and identify the viable parameter ranges\footnote{In \cite{Kallosh:2025rni}-\cite{Odintsov:2025wai} we refer to recent articles that propose to reconcile inflationary models with the latest CMB data.}. The analysis is deterministic and semi-analytical. For given model parameters we compute $n_s$, $r$, and $T_{re}$ and confront them directly with observations. A full Bayesian analysis is not attempted, since our goal is to check basic compatibility rather than perform parameter estimation. The study is therefore exploratory, providing a conceptual step toward a more detailed statistical treatment. Section~\ref{alfa} analyzes the generalized $\alpha$-attractor model. Section~\ref{staro} presents the $\alpha$-Starobinsky case. Section~\ref{con} summarizes the implications for the Planck–ACT–DESI interplay and clarifies that the upward shift in $n_s$ is a feature of CMB+DESI combinations rather than of CMB-only datasets.

\section{A generalized $\alpha$-attractor model}\label{alfa}
The $\alpha$-attractor models provide a simple class of inflationary potentials that give strong predictions for the spectral index $n_s$ and the tensor-to-scalar ratio $r$. A common generalization of the basic T-model ($p=2$) is given by \cite{Kallosh:2013yoa}
\begin{equation}
V(\phi) = V_0 \tanh^p\!\left(\frac{\phi}{\sqrt{6\alpha}M_{Pl}}\right),
\label{potanh}
\end{equation}
where $\alpha$ sets the curvature of the potential and $p$ modifies its shape. For $p=2$ this potential reproduces the original formulation of the T-models.

An alternative generalization, proposed in \cite{German:2021rin}, uses instead a $\sech$-function:
\begin{equation}
V(\phi) = V_0 \bigg(1 - \sech^p\!\left(\frac{\phi}{\sqrt{6\alpha}M_{Pl}}\right)\bigg).
\label{potsech}
\end{equation}
We refer to this form as the $\sech^p$ potential. Because $\sech(x)$ is positive definite for any real $x$, this potential is well defined for any positive $p$ (odd, even, or fractional), unlike \eqref{potanh} which restricts $p$ to even integers. For $p=2$ the two parameterizations coincide.

Expanding \eqref{potsech} near the minimum at $\phi=0$ gives
\begin{equation}
\frac{V}{V_0} = \frac{1}{2}p\!\left(\frac{\phi}{\sqrt{6\alpha}M_{Pl}}\right)^2 
- \frac{1}{24}p(2 + 3p)\!\left(\frac{\phi}{\sqrt{6\alpha}M_{Pl}}\right)^4 + \cdots,
\label{potorigins}
\end{equation}
so the leading term is always quadratic, regardless of $p$. This difference is important because for the $\tanh^p$ case the power $p$ enters also as a power in the terms of the expansion, giving stronger shape dependence.
To describe slow-roll dynamics we introduce the shorthand
\begin{equation}
Se \equiv \sech\!\left(\frac{\phi_k}{\sqrt{6\alpha}M_{Pl}}\right), \qquad
Si \equiv \sinh\!\left(\frac{\phi_k}{\sqrt{6\alpha}M_{Pl}}\right),
\label{definitions}
\end{equation}
where $\phi_k$ is the inflaton value at horizon crossing. The tensor-to-scalar ratio and spectral index are
\begin{align}
r &= \frac{4p^2}{3\alpha} \, \frac{Se^{2+2p} \, Si^2}{(1 - Se^p)^2}, 
\label{rsec} \\[6pt]
n_s &= 1 + \frac{p Se^p - p(p+1) Se^{p+2} Si^2}{3\alpha (1-Se^p)}
- \frac{p^2}{2\alpha}\, \frac{Se^{2+2p} Si^2}{(1-Se^p)^2},
\label{nssec}
\end{align}
obtained in the slow-roll approximation ($r=16\epsilon$, $n_s = 1+2\eta-6\epsilon$).
The number of $e$-folds from horizon exit until the end of inflation is
\begin{equation}
N_k(\phi_k,\phi_{end},\alpha,p) = -\frac{1}{M_{Pl}^2}\int_{\phi_k}^{\phi_{end}} \frac{V}{V'} \, d\phi,
\label{Nk}
\end{equation}
which for general $p$ involves hypergeometric functions. The end of inflation is defined by $\epsilon(\phi_{end})\simeq 1$.
For later use in reheating, we note the functions
\begin{equation}
f(\phi_k) = 1 - Se^p, \qquad 
f'(\phi_k) = \frac{p}{\sqrt{6\alpha}M_{Pl}}\, Se^{1+p} Si, \qquad
\gamma = \frac{p}{6\alpha},
\label{efes}
\end{equation}
where $\gamma$ is the dimensionless curvature of the potential at its minimum, defined by $V''(\phi=0) = \gamma V_0/M_{Pl}^2$ \cite{German:2025mzg}.

\subsection*{Choice of parameters}

The ranges of $(\alpha,p,\omega_{re})$ used in our analysis require justification. First, we note that ACT DR6 is compatible with Planck in CMB-only comparisons; the upward shift to higher $n_s$ emerges only when DESI BAO are combined with CMB datasets. In this context, it is still informative to explore model deformations that can accommodate the higher $n_s$ values preferred by CMB+DESI while remaining viable for CMB-only constraints. We consider $0<p<1$ because values smaller than unity shift predictions toward these higher $n_s$ values, though very small $p$ can lead to large $r$.
For this reason we highlight an intermediate case, $p=1/4$, as a representative scenario. We have ensured that within this range, $p$-values slightly different from 1/4 do not alter our conclusions. For the $\alpha$ parameter we explore values around $\alpha \sim \mathcal{O}(1)$, consistent with both theoretical motivations of $\alpha$-attractors and with Planck bounds on $r$. Also, $\alpha$ values close to 1 give rise to larger $n_s$ as illustrated in the figures. The reheating parameter is varied in the physically plausible interval $-1/3 \leq \omega_{re} \leq 1$, corresponding to matter-like reheating through to stiff post-inflationary expansion. These choices are consistent with previous analyses of reheating constraints in $\alpha$-attractors (see e.g.\, \cite{German:2024rmn}).

The numerical values of cosmological parameters used throughout are summarized in Table~\ref{values}. These depend only on well established measurements and set the physical normalization for the analysis.

\begin{table}[ht!]
\centering
\begin{tabular}{cccc}
\small
Parameter & Value & Parameter & Value \\
\hline \hline
$g_{re}$ & $106.75$ & $k_p$ & $0.05\, \mathrm{Mpc}^{-1}$ \\
$g_{s,re}$ & $106.75$ & $T_0$ & $2.7255\, \mathrm{K}$ \\
$A_s$ & $2.1 \times 10^{-9}$ & $a_{0}$ & $1$ \\
\hline \hline
\end{tabular}
\caption{Fixed parameter values used in numerical calculations.}
\label{values}
\end{table}

\begin{figure}[h!]
\centering 
\includegraphics[width=0.49\textwidth]{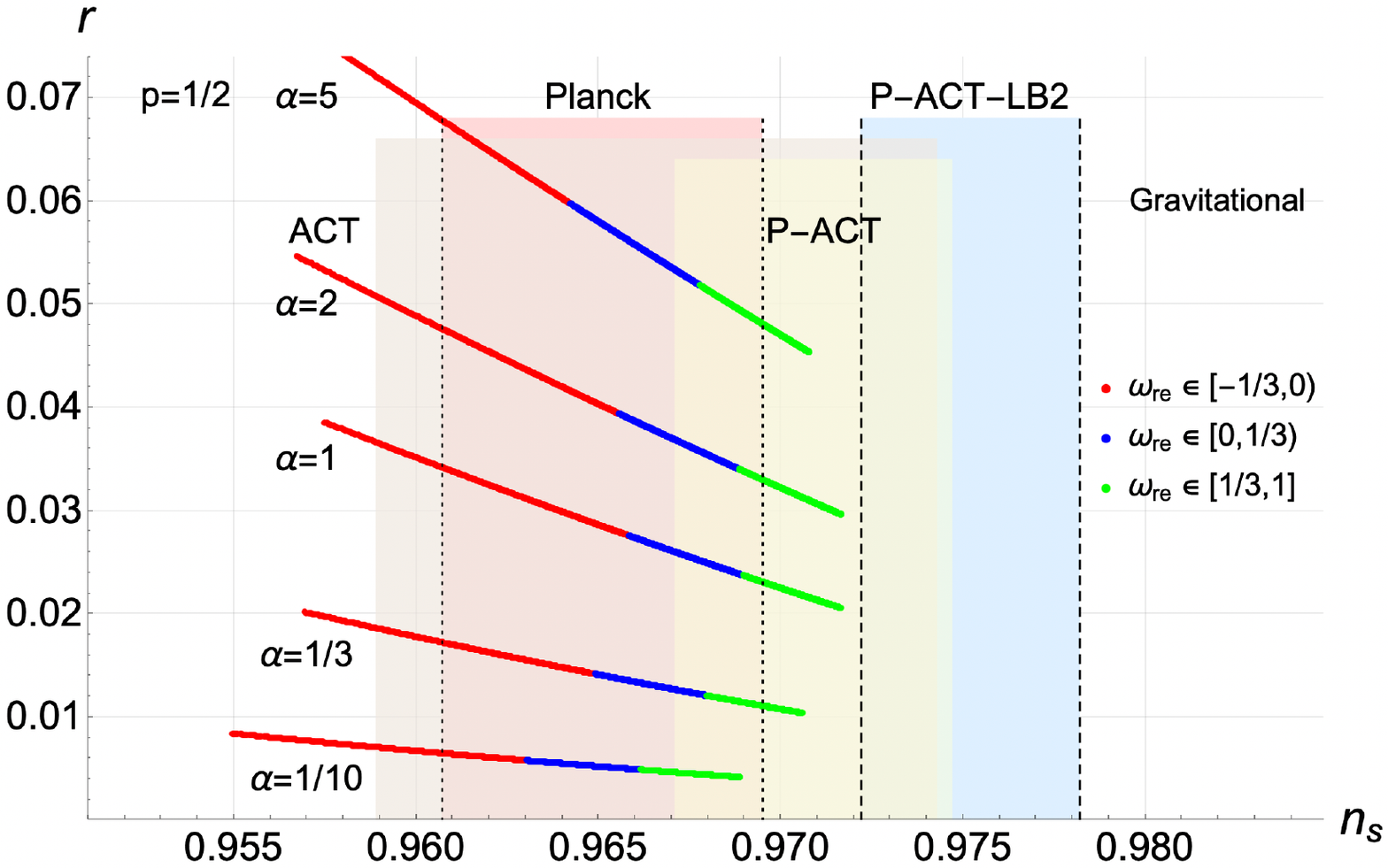}
\includegraphics[width=0.49\textwidth]{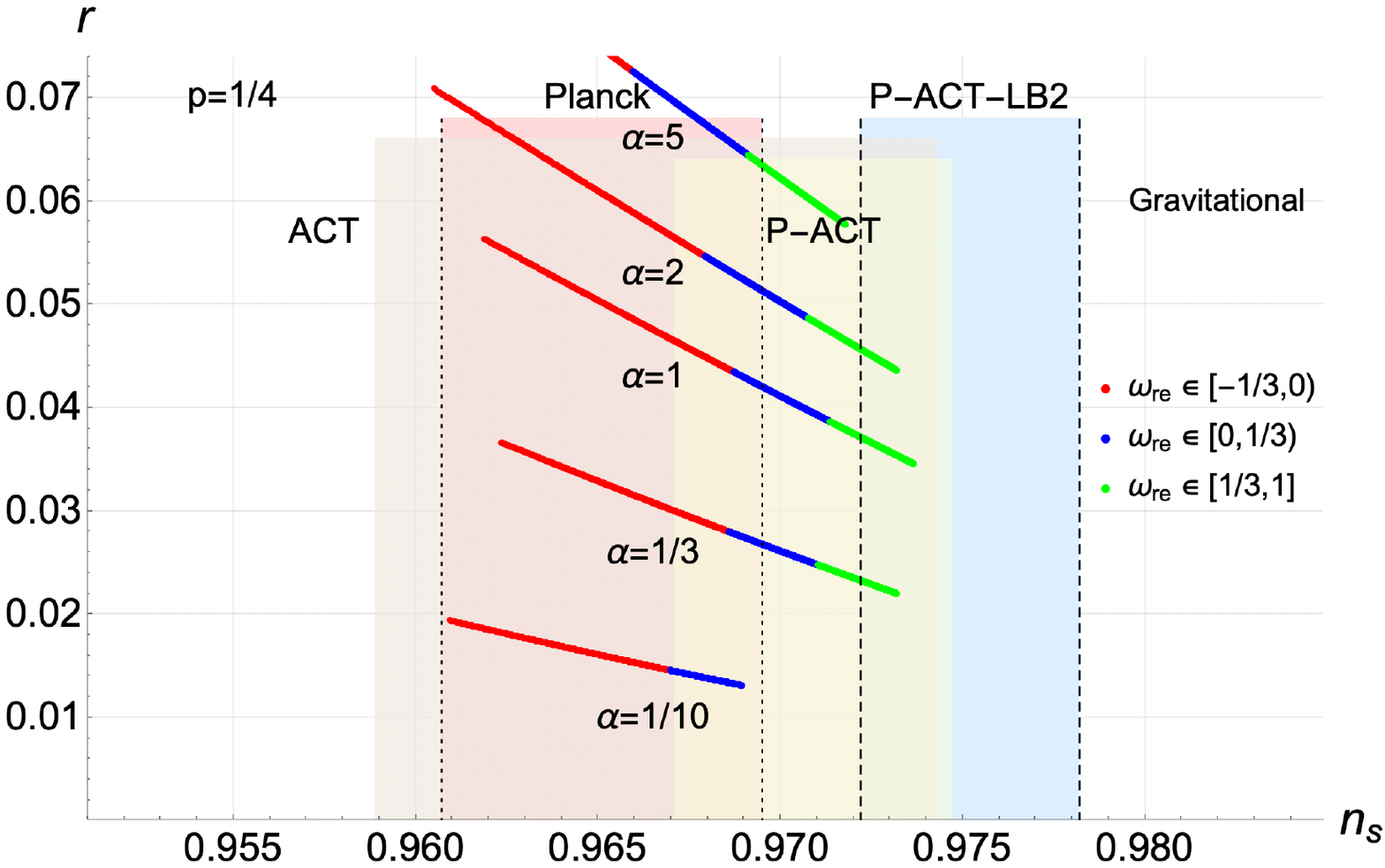}
\includegraphics[width=0.49\textwidth]{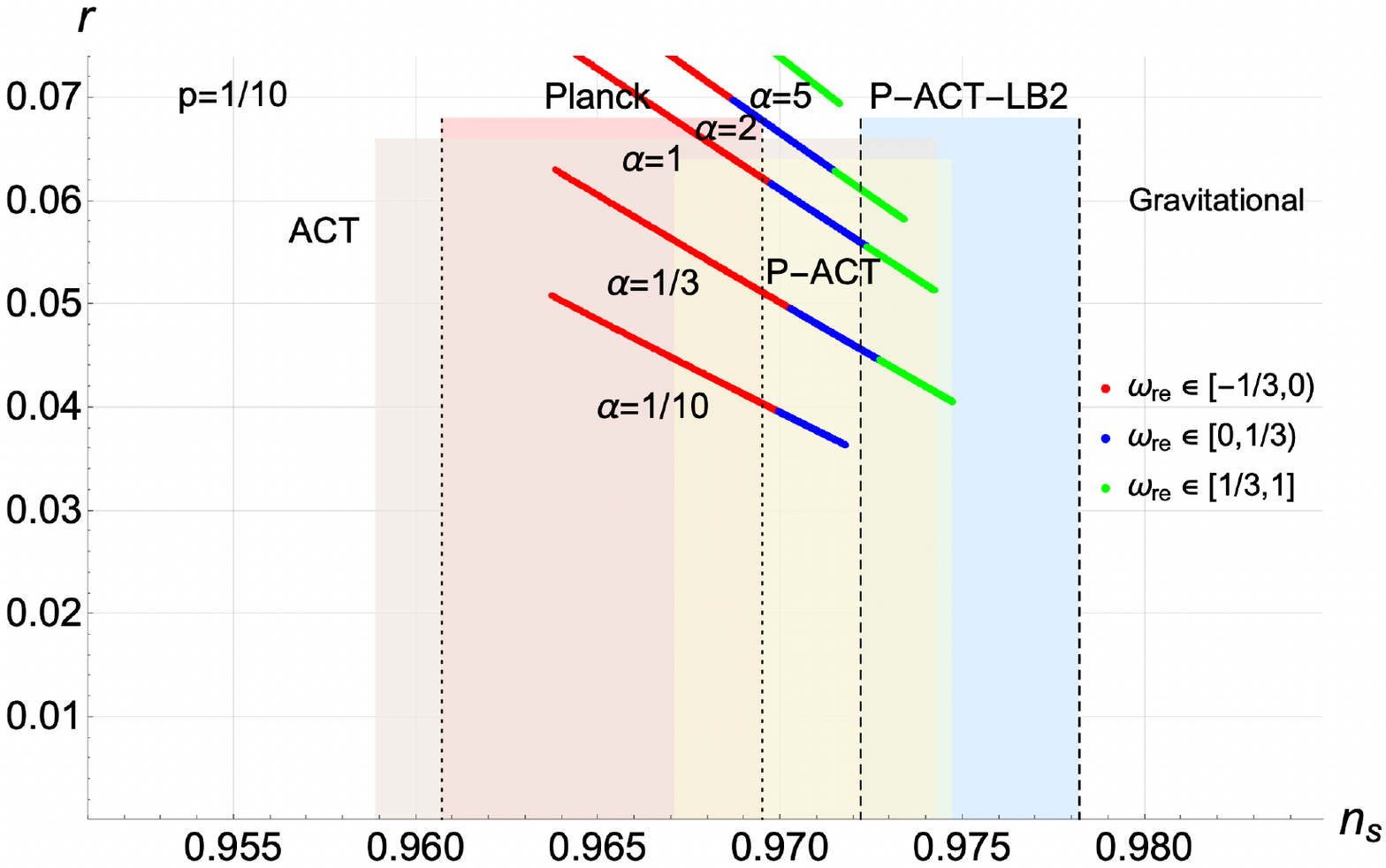}
\caption{Predictions of the generalized $\sech^p$ potential under gravitational reheating. 
Shown are trajectories in the $r$--$n_s$ plane for $p=1/2$, $p=1/4$, and $p=1/10$. 
The solid curves correspond to different values of $\alpha$, with colors denoting ranges of the reheating equation-of-state parameter: red for $-1/3 \leq \omega_{re} < 0$, blue for $0 \leq \omega_{re} < 1/3$, and green for $1/3 \leq \omega_{re} \leq 1$. 
The shaded regions indicate the $1\sigma$ intervals favored by current observations \cite{ACT:2025fju}: Planck (light red, $n_s=0.9651\pm0.0044$), ACT DR6 (light brown, $n_s=0.9666\pm0.0077$), P+ACT (light yellow, $n_s=0.9709\pm0.0038$), and P-ACT-LB2 (light blue, $n_s=0.9752\pm0.0030$). 
Decreasing $p$ shifts the curves toward the P-ACT-LB2-preferred region but simultaneously increases $r$.  For very small $p$ the predictions exceed the current upper bound on $r$.}
\label{sechregiones}
\end{figure}

With these definitions and parameter ranges, the next step is to apply Eq.~(\ref{principal}) to compute $\phi_k$ and thereby obtain predictions for reheating temperatures in the different decay channels studied here (gravitational, fermionic, and scalar). The results are shown in Figure~\ref{sechregiones} and discussed in the following subsections.
\subsection{Gravitational decay}\label{gravitational}

The perturbative reheating temperature for purely gravitational inflaton decay is \cite{German:2025mzg}
\begin{equation}
\label{Tregra}
T_{re}^{\mathrm{(grav)}} \simeq
\left( \frac{135\, A_s^3 \pi^2 f'(\phi_k)^6 M_{Pl}^6 \gamma^3}{32\, g_{re}\, f(\phi_k)^9} \right)^{1/4} M_{Pl}.
\end{equation}
Substituting \eqref{Tregra} into Eq.~(\ref{principal}) allows us to solve for $\phi_k$, given the model parameters $(\alpha,p)$ and the reheating equation-of-state parameter $\omega_{re}$. The physical range $-1/3 \leq \omega_{re} \leq 1$ is sampled in three intervals, displayed with distinct colors in the figures: red ($-1/3 \leq \omega_{re}<0$), blue ($0 \leq \omega_{re}<1/3$), and green ($1/3 \leq \omega_{re} \leq 1$). 

\begin{figure}[h]
\centering
\includegraphics[width=4.5in]{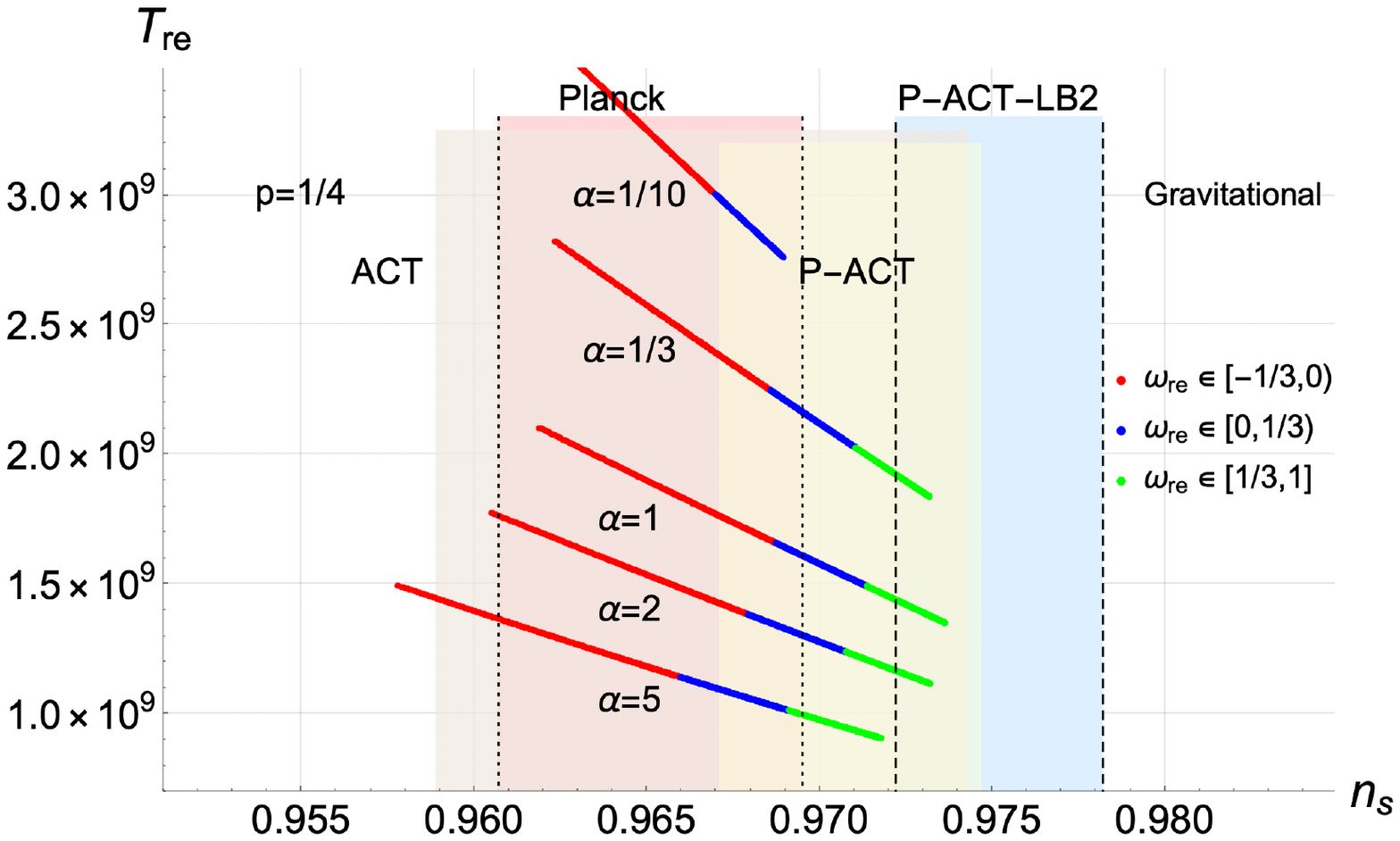}
\caption{Reheating predictions for the $\sech^p$ potential with $p=1/4$ assuming gravitational decay. 
The plot shows $T_{re}$ as a function of $n_s$, obtained by solving Eq.~(\ref{principal}) with $T_{re}$ given by Eq.~(\ref{Tregra}). 
Curves correspond to different $\alpha$ values, color-coded by $\omega_{re}$ as in Fig.~\ref{sechregiones}. 
For $\alpha \sim 1$, the trajectories approach the P-ACT - LB2 region, though only for $\omega_{re}$ close to the stiff limit $\omega_{re}\to1$ (green color).}
\label{Sech_Grav}
\end{figure}

Figure~\ref{sechregiones} shows $r$--$n_s$ predictions of the $\sech^p$ model for $p=1/2,\,1/4,\,1/10$. As $p$ decreases below unity, the curves move toward the regions favored by ACT DR6 and P--ACT--LB2 in CMB+DESI combinations, but simultaneously the tensor-to-scalar ratio $r$ grows. For sufficiently small $p$, the resulting $r$ exceeds the current observational upper limits. To illustrate this balance and for concreteness, in the following we focus on the intermediate case $p=1/4$. The corresponding $T_{re}$--$n_s$ predictions for gravitational decay are presented in Fig.~\ref{Sech_Grav}.

\subsection{Decay into fermions}\label{fermions}

\begin{figure}[h]
\centering
\includegraphics[width=4.5in]{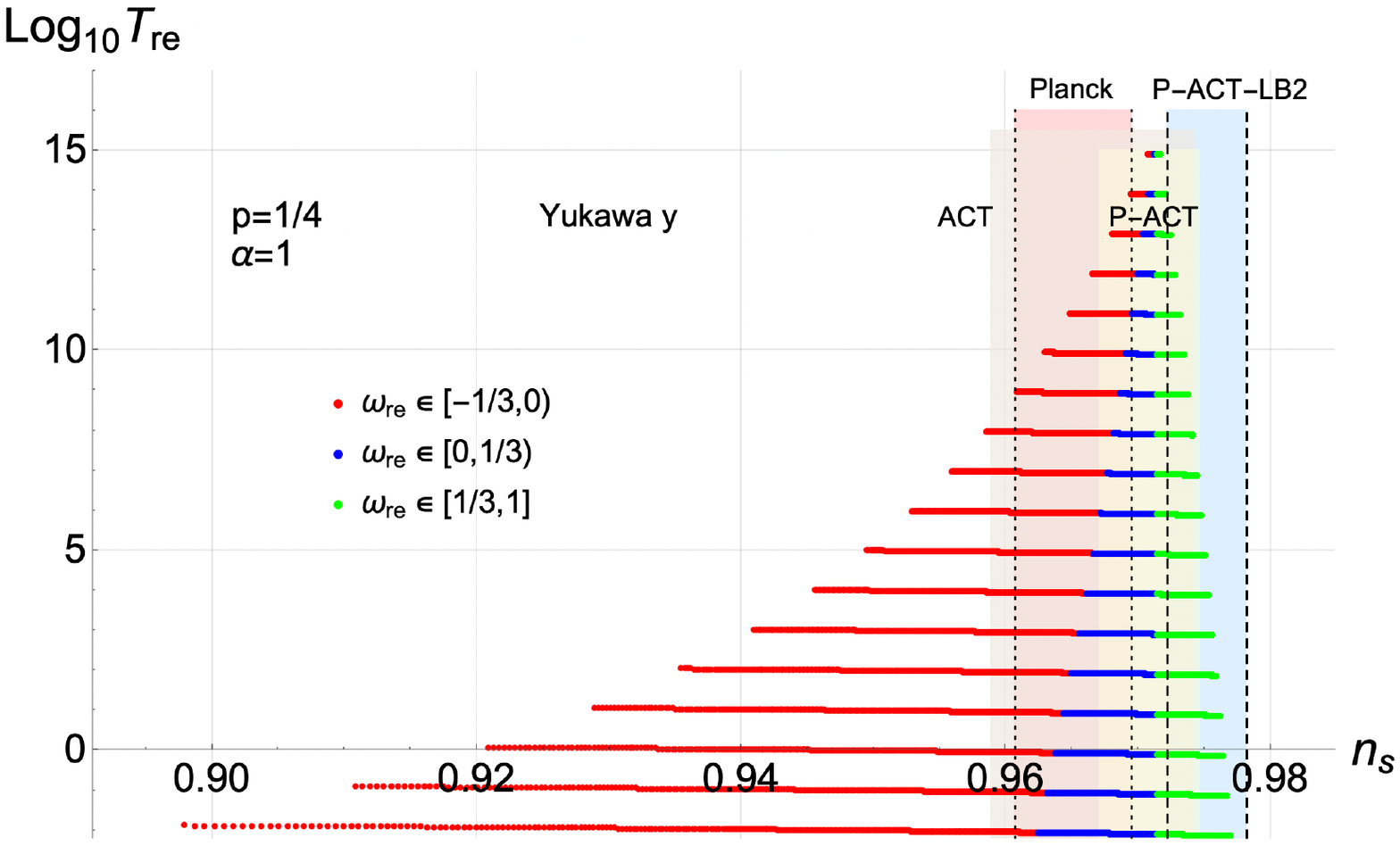}
\caption{Reheating predictions for the $\sech^p$ potential with $p=1/4$ assuming Yukawa (fermionic) decay. 
The plot shows $\log_{10} T_{re}$ versus $n_s$ obtained from Eq.~(\ref{Trey}). 
The Yukawa coupling $y$ is varied from $10^{-17}$ to unity. 
Larger $y$ values yield higher reheating temperatures and reduce the allowed range of $n_s$. 
For $\omega_{re}=1/3$, the dependence on $T_{re}$ cancels, and the transition between the blue and green segments occurs at a fixed $n_s$, independent of $y$.}
\label{LogTreYuka}
\end{figure}

If the inflaton decays via a Yukawa coupling $y$ to fermions, the reheating temperature is \cite{German:2025mzg}
\begin{equation}
\label{Trey}
T_{re}^{(y)} \simeq |y| 
\left( \frac{135\, A_s\, f'(\phi_k)^2 M_{Pl}^2 \gamma}{8 \pi^2 g_{re}\, f(\phi_k)^3} \right)^{1/4} M_{Pl}.
\end{equation}
After solving Eq.~(\ref{principal}) for $\phi_k$, we compute $T_{re}^{(y)}$ from \eqref{Trey}. Figure~\ref{LogTreYuka} shows $\log_{10}T_{re}$ versus $n_s$, with $y$ varied between $10^{-17}$ and 1. Larger values of $y$ lead to higher reheating temperatures, but also reduce the allowed range of $n_s$. A specific feature appears at $\omega_{re}=1/3$ showing that the location of the blue–green transition point along the $n_s$ axis is independent of $y$, for fixed $\alpha$. This follows directly from Eq.~(\ref{principal}), where all explicit dependence on $T_{re}$ cancels when $\omega_{re}=1/3$.

\subsection{Decay into scalars}\label{scalars}

\begin{figure}[h]
\centering
\includegraphics[width=4.5in]{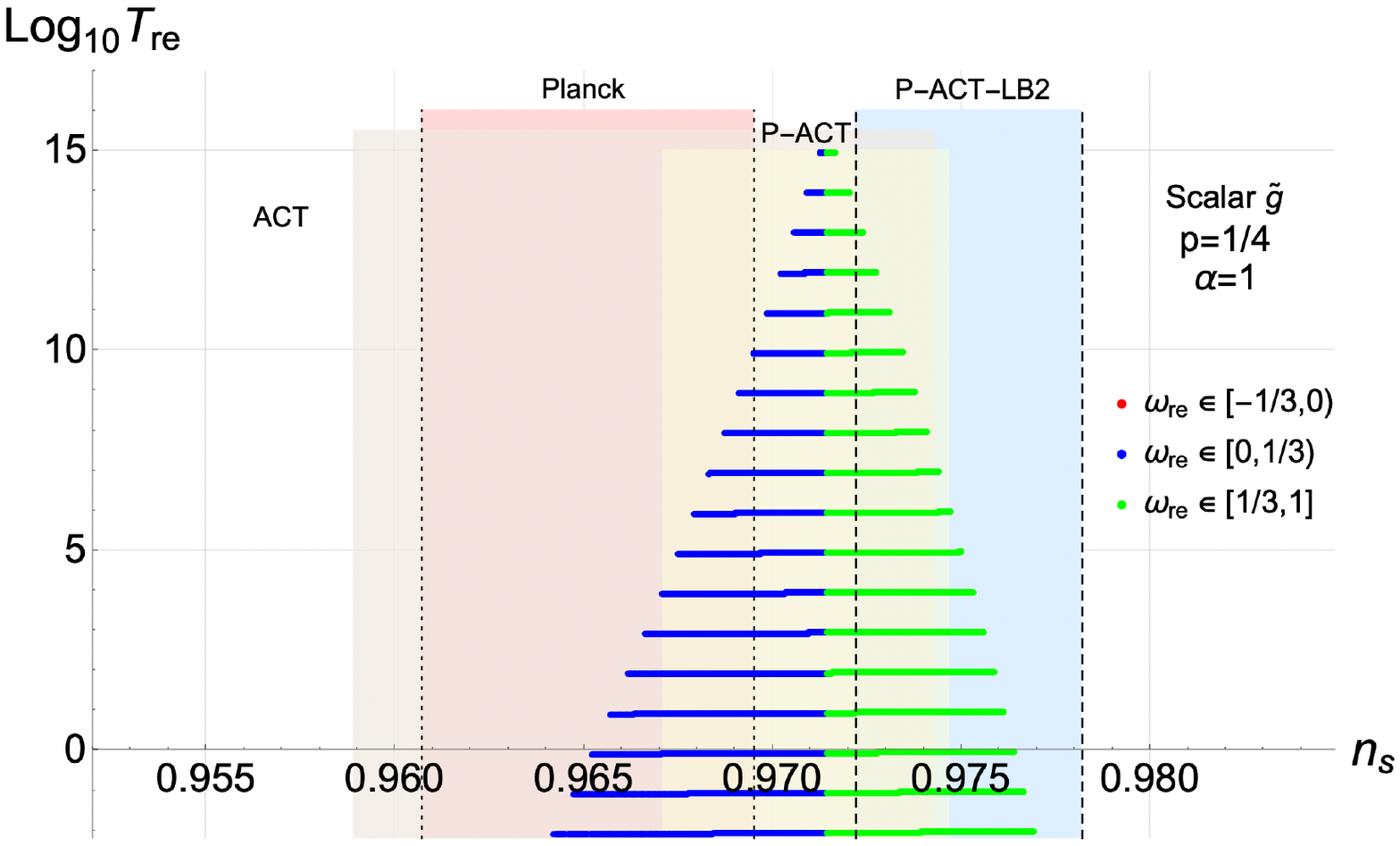}
\caption{Reheating predictions for the $\sech^p$ potential with $p=1/4$ assuming scalar decay. 
The vertical axis shows $\log_{10} T_{re}$ computed from Eq.~(\ref{Treg}), with the effective coupling $\tilde{g}=g/M_{Pl}$ varied between $10^{-23}$ and $10^{-5}$. 
Increasing $\tilde{g}$ raises the reheating temperature and narrows the allowed $n_s$ interval, more strongly than in the Yukawa case (Fig.~\ref{LogTreYuka}). 
As in other decay scenarios, the blue–green transition corresponds to $\omega_{re}=1/3$ and lies at fixed $n_s$ for a given $\alpha$.}
\label{LogTreScalar}
\end{figure}

For perturbative decay into scalar fields with coupling $g$, it is convenient to define the dimensionless parameter $\tilde g \equiv g/M_{Pl}$. The reheating temperature in this case is \cite{German:2025mzg}
\begin{equation}
\label{Treg}
T_{re}^{(g)} \simeq |\tilde g| 
\left( \frac{15\, f(\phi_k)^3}{128\, A_s \pi^6 f'(\phi_k)^2 M_{Pl}^2 g_{re}\, \gamma} \right)^{1/4} M_{Pl}.
\end{equation}
Figure~\ref{LogTreScalar} displays $\log_{10}T_{re}$ versus $n_s$, with $\tilde g$ scanned in the range $10^{-23}\leq \tilde g \leq 10^{-5}$. Increasing $\tilde g$ raises the reheating temperature and, more significantly than in the Yukawa case, narrows the admissible interval of $n_s$. As before, when $\omega_{re}=1/3$ the transition between blue and green segments occurs at a fixed $n_s$ for a given $\alpha$, reflecting the cancellation of $T_{re}$ in Eq.~(\ref{principal}).

In the previous parameter scans we include extremely small couplings, e.g., $y \sim 10^{-17}$ and $\tilde g \sim 10^{-23}$. While such values are mathematically allowed and useful for mapping out the full dependence of $T_{re}$ and $n_s$, they may be considered non-generic from a particle-physics perspective and could require additional model-building motivation; our use here is phenomenological.

\section{The $\alpha$-Starobinsky model}\label{staro}
The Starobinsky model $V(\phi) \propto (1 - e^{-\sqrt{2/3}\,\phi/M_{Pl}})^2$ \cite{Starobinsky:1980te} is well known for predicting $n_s$ and $r$ values in excellent agreement with Planck data. The $\alpha$-Starobinsky generalization \cite{Ellis:2013nxa, Kallosh:2013yoa, Ellis:2019bmm} modifies the exponential term through a free parameter $\alpha$:  
\begin{equation}
V(\phi) = V_0 \left(1 - e^{-\sqrt{\frac{2}{3\alpha}} \frac{\phi}{M_{Pl}}} \right)^2,
\label{Staro_pot}
\end{equation}
For $\alpha=1$ this reduces to the original Starobinsky potential, while varying $\alpha$ controls the flattening of the potential at large $\phi$, thereby shifting predictions for $r$ and (to a lesser degree) $n_s$. This makes it a natural test case for examining the higher $n_s$ values that emerge in CMB+DESI DR2 combinations, and we keep this in mind when scanning $\alpha$ to reach larger $n_s$.

The slow-roll parameters can be written in terms of  
\begin{equation}
u \equiv e^{-\sqrt{\frac{2}{3\alpha}} \frac{\phi_k}{M_{Pl}}} ,
\end{equation}
leading to 
\begin{align}
r &= \frac{64}{3\alpha} \, \frac{u^2}{(1-u)^2}, \label{Staro_r}\\[6pt]
n_s &= 1 - \frac{8}{3\alpha} \, \frac{u(1+u)}{(1-u)^2}. \label{Staro_ns}
\end{align}
The number of $e$-folds between horizon exit and the end of inflation is
\begin{equation}
N_k(\phi_k, \phi_{end}, \alpha) 
= -\frac{1}{M_{Pl}^2} \int_{\phi_k}^{\phi_{end}} \frac{V}{V'}\, d\phi = \frac{3\alpha}{4} \left(\frac{1}{u}-\frac{1}{u_{e}}+\ln\left(\frac{u}{u_{e}}\right)  \right),
\label{Staro_Nk}
\end{equation}
where $u_e \equiv e^{-\sqrt{\frac{2}{3\alpha}} \frac{\phi_{end}}{M_{Pl}}}$ with $\phi_{end}$ defined by $\epsilon(\phi_{end}) \simeq 1$. These relations allow us to compute all needed quantities for reheating.

For Eq.~(\ref{principal}), the relevant potential functions take the form
\begin{equation}
f(\phi_k) = (1-u)^2, \qquad
f'(\phi_k) = \sqrt{\frac{2}{3\alpha}}\, \frac{2}{M_{Pl}}\, u (1-u), \qquad
\gamma = \frac{4}{3\alpha}.
\label{Staro_fdefs}
\end{equation}
With these ingredients, the same three decay scenarios as in Sec.~\ref{alfa} can be analyzed.

\subsection{Gravitational decay}

\begin{figure}[h]
\centering
\begin{tabular}{cc}
\includegraphics[width=3.2in]{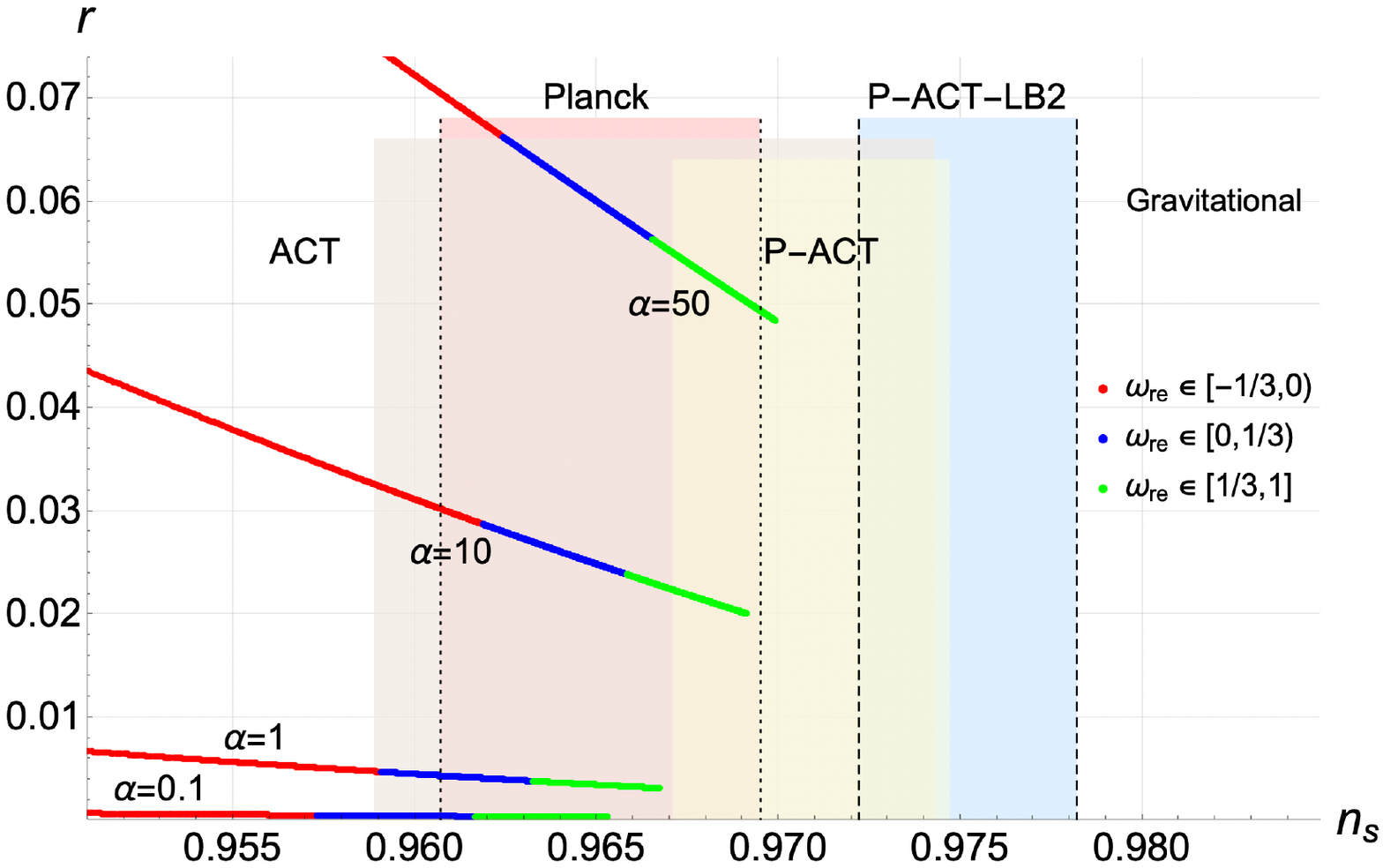} &
\includegraphics[width=3.2in]{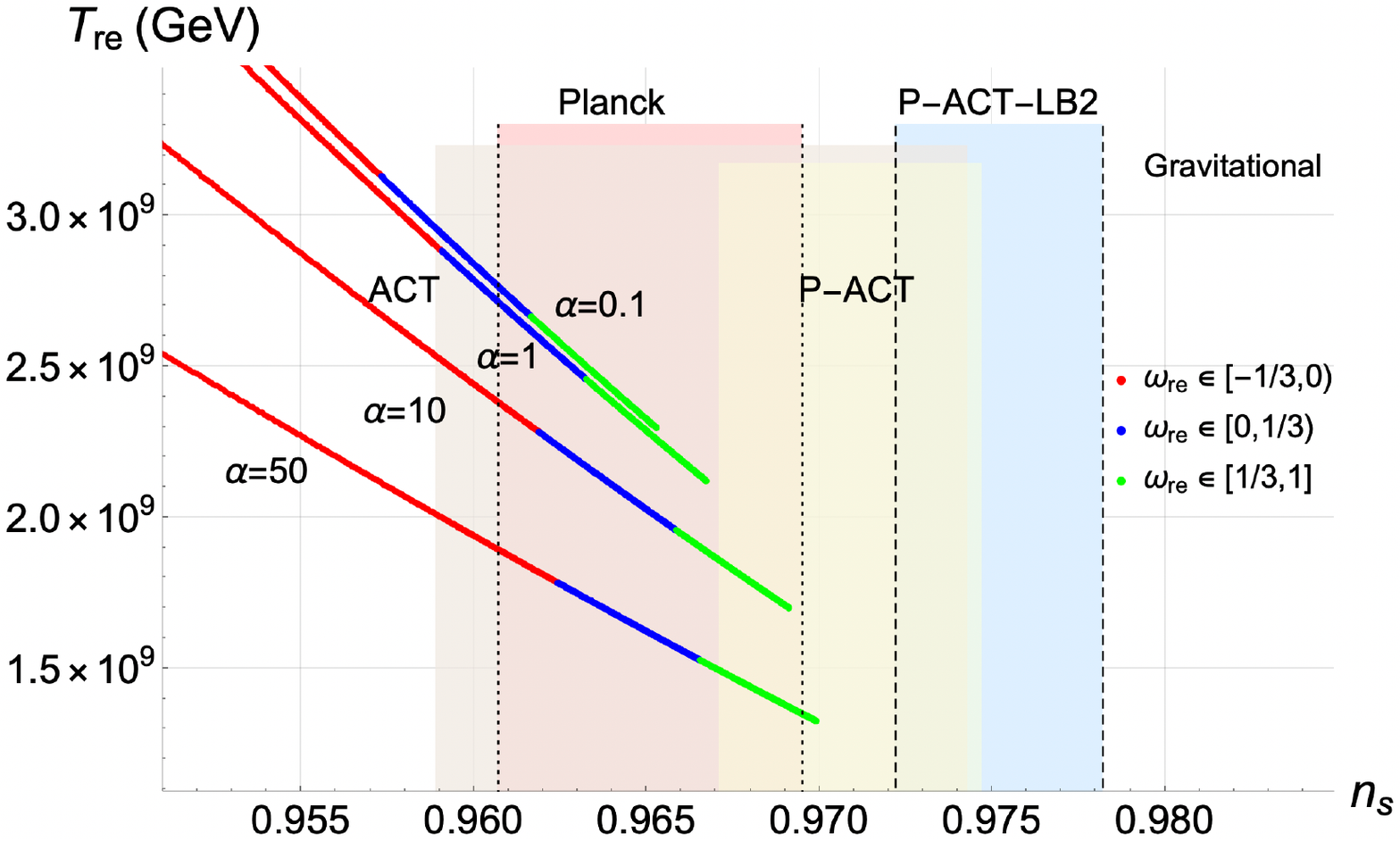}
\end{tabular}
\caption{Predictions of the $\alpha$-Starobinsky potential under gravitational decay. 
Left: trajectories in the $r$--$n_s$ plane. 
Right: corresponding reheating predictions in the $T_{re}$--$n_s$ plane, obtained by solving Eq.~(\ref{principal}) with $T_{re}$ from Eq.~(\ref{Tregra}). 
Curves represent different $\alpha$ values, color-coded for $\omega_{re}$ as in Fig.~\ref{sechregiones}. 
Larger $\alpha$ shifts the curves toward the P-ACT-LB2-preferred $n_s$ region but simultaneously increases $r$. 
For $\alpha \gtrsim 10$, $r$ exceeds the current upper bounds, setting an approximate upper limit for this scenario.}
\label{Staro_Grav}
\end{figure}

The gravitational reheating temperature follows from Eq.~(\ref{Tregra}), evaluated with functions \eqref{Staro_fdefs}. Figure~\ref{Staro_Grav} shows the resulting $r$--$n_s$ and $T_{re}$--$n_s$ predictions. Increasing $\alpha$ shifts the curves toward higher $n_s$ values that can overlap with the regions preferred when ACT data are combined with DESI DR2, but simultaneously increases $r$. Values $\alpha \gtrsim 10$ generally produce $r$ above current observational bounds, suggesting an approximate upper limit for viable $\alpha$.

\subsection{Decay into fermions}

\begin{figure}[h]
\centering
\begin{tabular}{cc}
\includegraphics[width=3.2in]{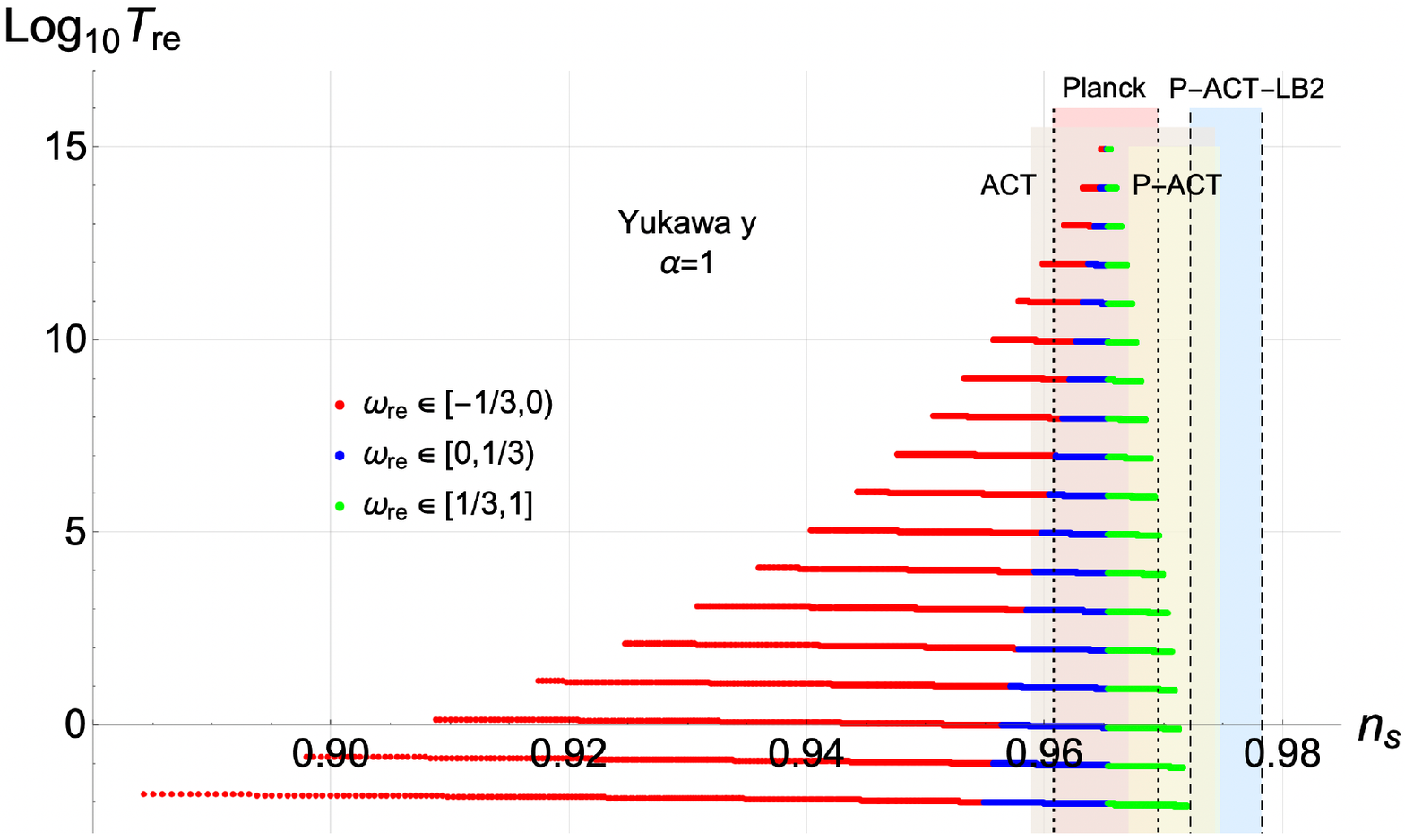} &
\includegraphics[width=3.2in]{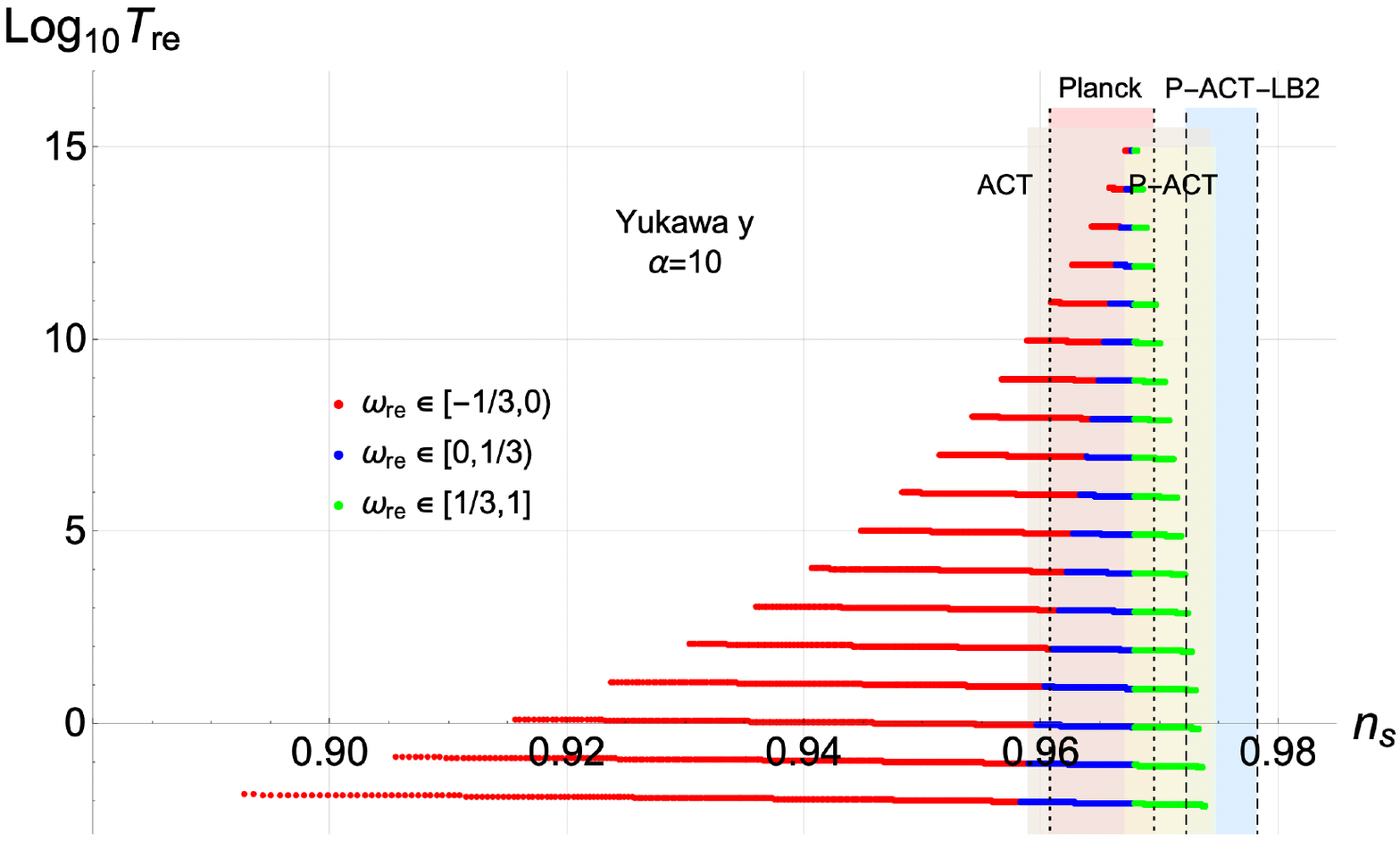}
\end{tabular}
\caption{Reheating predictions for the $\alpha$-Starobinsky potential with Yukawa (fermionic) decay. 
The plots show $\log_{10} T_{re}$ versus $n_s$, obtained from Eq.~(\ref{Trey}), for two representative values of $\alpha$: $\alpha=1$ (left) and $\alpha=10$ (right). 
The Yukawa coupling $y$ is varied from $10^{-17}$ to 1. 
Larger $y$ gives higher $T_{re}$ values and reduces the accessible $n_s$ interval. 
For $\alpha=10$ the curve approaches the P-ACT-LB2 region but lies near the observational upper bound on $r$. 
As in other reheating channels, the transition between blue and green segments corresponds to $\omega_{re}=1/3$, where $T_{re}$ drops out of Eq.~(\ref{principal}).}
\label{Staro_Yuka}
\end{figure}

For a Yukawa coupling $y$, the reheating temperature is given by Eq.~(\ref{Trey}). Figure~\ref{Staro_Yuka} shows $\log_{10}T_{re}$ against $n_s$ for $\alpha=1$ (left panel) and $\alpha=10$ (right panel), with $y$ scanned between $10^{-17}$ and 1. As in the $\sech^p$ case, larger $y$ raises $T_{re}$ and narrows the $n_s$ interval. For $\alpha=10$, the trajectories approach the $n_s$ region favored by CMB+DESI (including P-ACT-LB2) but also lie close to the current upper bound on $r$. The blue--green transition, corresponding to $\omega_{re}=1/3$, appears at fixed $n_s$ and is independent of $y$.

\subsection{Decay into scalars}

\begin{figure}[h]
\centering
\begin{tabular}{cc}
\includegraphics[width=3.2in]{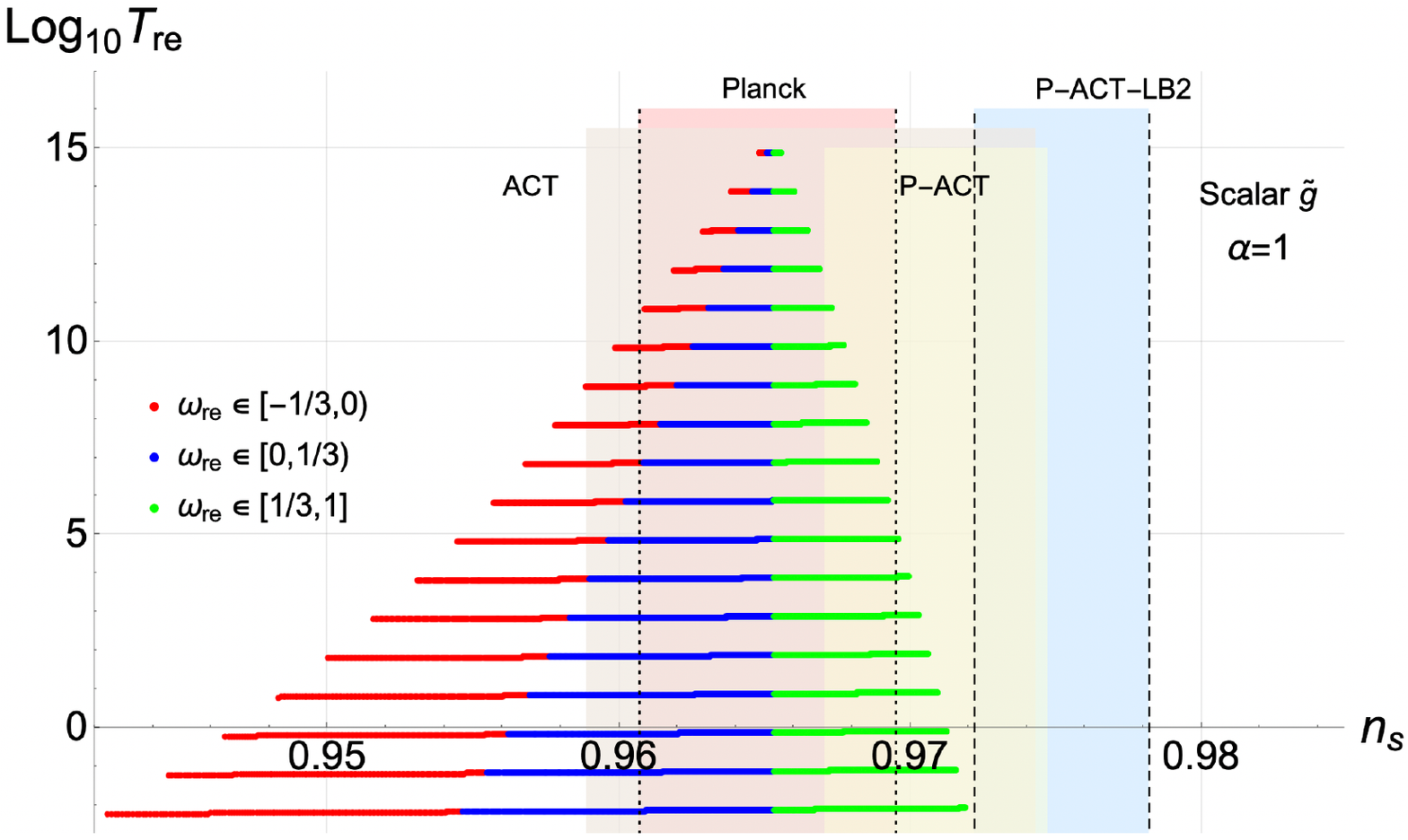} &
\includegraphics[width=3.2in]{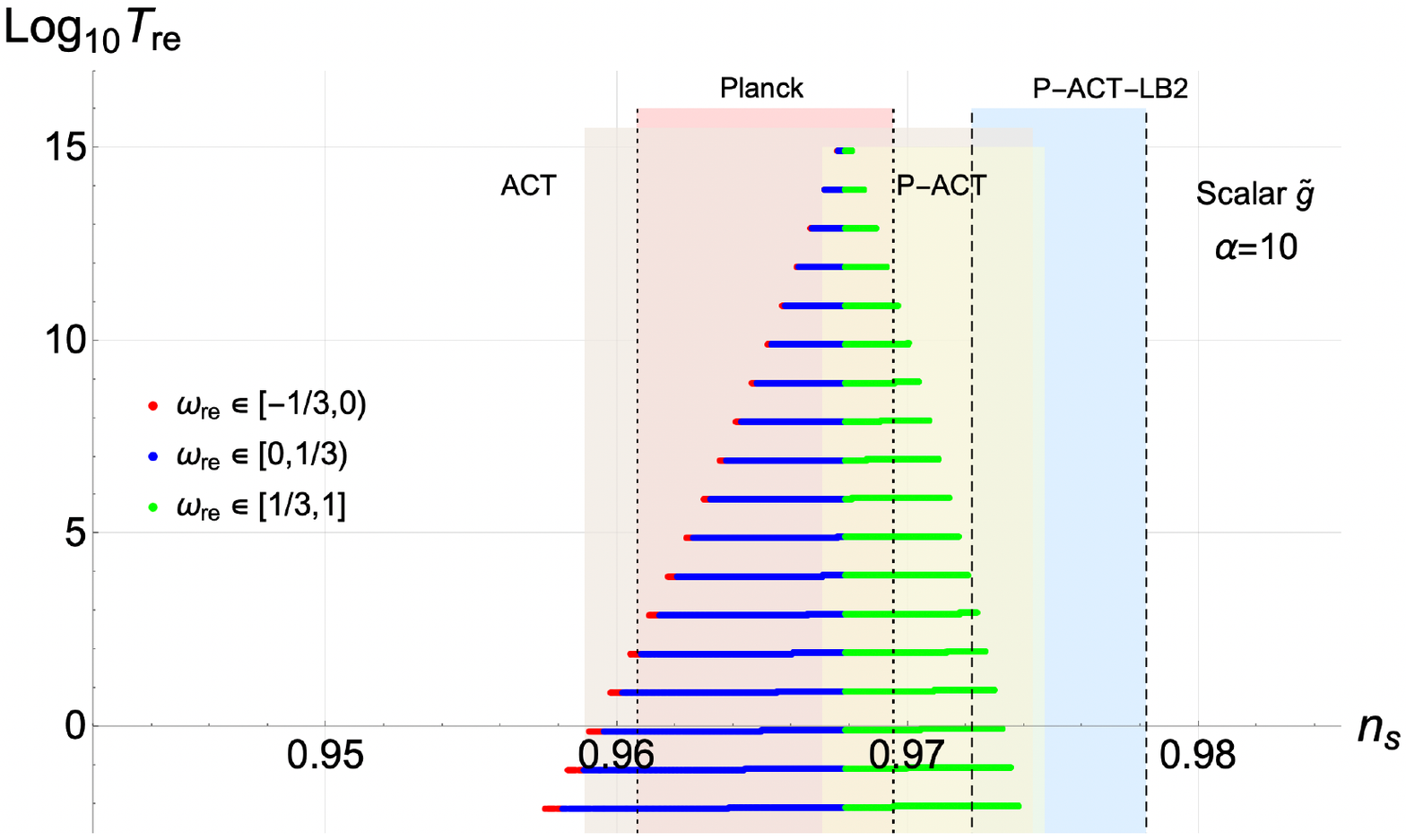}
\end{tabular}
\caption{Reheating predictions for the $\alpha$-Starobinsky potential with scalar decay. 
The vertical axis shows $\log_{10} T_{re}$ computed from Eq.~(\ref{Treg}), with $\tilde{g}=g/M_{Pl}$ varied between $10^{-23}$ and $10^{-5}$. 
Results are shown for $\alpha=1$ (left) and $\alpha=10$ (right). 
Increasing $\tilde{g}$ raises $T_{re}$ and narrows the allowed range of $n_s$. 
For $\alpha=10$ the trajectories move closer to the P-ACT-LB2 region but approach the observational upper limits on $r$. 
As in the Yukawa case, the blue–green transition corresponds to $\omega_{re}=1/3$, with fixed $n_s$ for a given $\alpha$.}
\label{Staro_Scalar}
\end{figure}

For scalar decay with coupling $g$, expressed as $\tilde g = g/M_{Pl}$, the reheating temperature follows from Eq.~(\ref{Treg}). Results are shown in Fig.~\ref{Staro_Scalar} for $\alpha=1$ and $\alpha=10$. As $\tilde g$ increases (from $10^{-23}$ to $10^{-5}$), $T_{re}$ grows and the allowed $n_s$ interval becomes narrower, with a stronger effect than in the Yukawa case. For larger $\alpha$, the trajectories intersect the higher $n_s$ values preferred by CMB+DESI combinations, but at the price of larger $r$, again indicating tension with current observational bounds.

\section{Discussion and Conclusions}\label{con}

In this work, we examined the interplay between inflation and the subsequent reheating phase in two classes of single-field inflationary models: a generalized $\alpha$-attractor model based on a hyperbolic secant potential, and the $\alpha$-Starobinsky model. Our goal was to assess their compatibility with recent cosmological observations, including data from Planck, the Atacama Cosmology Telescope (ACT DR6), South Pole Telescope (SPT-3G), and baryon acoustic oscillation (BAO) measurements from DESI DR2.

Using a semi-analytical approach based on a relation between inflationary dynamics and reheating, we computed the inflaton field value at horizon crossing $\phi_k$ and, from it, derived predictions for the scalar spectral index $n_s$, the tensor-to-scalar ratio $r$, and the reheating temperature $T_{re}$. We considered three types of inflaton decay channels—gravitational, Yukawa (fermionic), and scalar—and explored how these affect the predictions in the $r$--$n_s$ and $T_{re}$--$n_s$ planes.

For the $\sech^p$ potential of Section \ref{alfa}, we found that decreasing $p$ below 1 shifts the model predictions toward the higher $n_s$ values favored by combinations that include DESI DR2 (e.g., Planck+DESI, ACT DR6+DESI, SPT-3G+DESI), as well as the P–ACT–LB2 combination reported by ACT. However, very small values of $p$ also lead to large values of $r$, which can exceed current observational bounds. The intermediate case $p=1/4$ offers a good compromise and was used in most of our quantitative analysis.

For the $\alpha$-Starobinsky model, larger values of $\alpha$ shift the model predictions toward these higher-$n_s$ regions seen in CMB+DESI combinations. However, for $\alpha \gtrsim 10$, the predicted tensor-to-scalar ratio $r$ becomes too large, conflicting with observational bounds. We thus identify $\alpha \approx 10$ as an approximate upper bound for this model under gravitational reheating.

In all decay channels, the transition points between the blue and green segments in the $T_{re}$–$n_s$ plots remain fixed along the $n_s$ axis, for $\alpha$ fixed. This behavior corresponds to $\omega_{re} = 1/3$, for which the $T_{re}$ dependence drops out of the main equation, Eq.~(\ref{principal}). The scalar decay channel generally produces a narrower range of $n_s$ values than the Yukawa channel.

A clarification regarding the scalar spectral index is in order. ACT DR6, considered as a CMB-only dataset, is fully compatible with Planck and does not by itself prefer a higher $n_s$. The upward shift to $n_s \simeq 0.975$ emerges when CMB datasets (Planck, ACT DR6, or SPT-3G) are combined with DESI DR2 BAO. The same trend is present in the P–ACT–LB2 combination reported by ACT. In particular, the value $n_s = 0.9752 \pm 0.0030$ arises in combinations that include DESI DR2 \cite{ACT:2025fju}. This central value is higher than Planck’s $n_s = 0.9651 \pm 0.0044$ by about $\Delta n_s \simeq 0.01$, corresponding to a discrepancy exceeding $2\sigma$ depending on priors and analysis choices. The origin of this difference is still under investigation and may reflect residual systematics, methodological differences, or physics beyond $\Lambda$CDM.

In conclusion, both the generalized $\alpha$-attractor and $\alpha$-Starobinsky models can be consistent with current observations under specific parameter choices and decay scenarios. Their compatibility with the higher-$n_s$ regions seen in CMB+DESI combinations (including P–ACT–LB2 and SPT-3G+DESI) often comes with increased $r$, which can strain current bounds. Further refinement of CMB measurements—particularly of $n_s$ and $r$—and improved modeling of reheating and decay channels will be crucial for testing the viability of these inflationary scenarios.
\acknowledgments

We would like to thank DGAPA-PAPIIT-UNAM grant No. IN110325 {\it Estudios en cosmolog\'ia inflacionaria, agujeros negros primordiales y energ\'ia oscura}, DGAPA-PAPIIT-UNAM grant No. IG102123: {\it Laboratorio de Modelos y datos para proyectos de investigación científica: Censos Astrofísicos} and SECIHTI grant CBF 2023-2024-162. We would like to thank the anonymous referee for clarifying remarks.

\newpage

\end{document}